\documentclass[aps,prd,twocolumn,preprintnumbers,amsmath,amssymb,nofootinbib,superscriptaddress]{revtex4-2}

\usepackage{graphicx}
\usepackage{booktabs}
\usepackage{longtable}
\usepackage{bm}
\usepackage{multirow}
\usepackage{hyperref}
\usepackage{xcolor}

\begin{document}
	
	\title{Clarifying the puzzling mass shift of the $\psi(4160)$ via a reanalysis of $R$-value data with unquenched charmonium spectroscopy}
	
	\author{Tian-Cai Peng}
	\email{pengtc20@lzu.edu.cn}
	\affiliation{School of Physical Science and Technology, Lanzhou University, Lanzhou 730000, China}
	\affiliation{Lanzhou Center for Theoretical Physics, Key Laboratory of Theoretical Physics of Gansu Province, Key Laboratory of Quantum Theory and Applications of MoE, Gansu Provincial Research Center for Basic Disciplines of Quantum Physics, Lanzhou University, Lanzhou 730000, China}
	\affiliation{Research Center for Hadron and CSR Physics, Lanzhou University and Institute of Modern Physics of CAS,	Lanzhou 730000, China}
	
	\author{Xiang Liu}
	\email{xiangliu@lzu.edu.cn}
	\affiliation{School of Physical Science and Technology, Lanzhou University, Lanzhou 730000, China}
	\affiliation{Lanzhou Center for Theoretical Physics, Key Laboratory of Theoretical Physics of Gansu Province, Key Laboratory of Quantum Theory and Applications of MoE, Gansu Provincial Research Center for Basic Disciplines of Quantum Physics, Lanzhou University, Lanzhou 730000, China}
	\affiliation{Research Center for Hadron and CSR Physics, Lanzhou University and Institute of Modern Physics of CAS,	Lanzhou 730000, China}
	\affiliation{MoE Frontiers Science Center for Rare Isotopes, Lanzhou University, Lanzhou 730000, China}

	\date{\today}
	
	\begin{abstract}
		The long-standing upward shift of the extracted $\psi(4160)$ mass, from about $4.16$~GeV to $4.19$~GeV in later analyses, remains a puzzling issue in charmonium spectroscopy. In our previous study, this problem was investigated through the $B^+\to K^+\mu^+\mu^-$ process within an unquenched charmonium framework, where the lower-mass $\psi(4160)$ assignment was found to be compatible with the data. Here we revisit the BESII $R$-value data, which played an important role in the historical extraction of the higher $\psi(4160)$ mass, and provide an independent examination. In contrast to the conventional quenched picture with $\psi(4040)$, $\psi(4160)$, and $\psi(4415)$, the unquenched vector-charmonium spectrum contains six states: $\psi(4040)$, $\psi(4160)$, $\psi(4220)$, $\psi(4380)$, $\psi(4415)$, and $\psi(4500)$. Including these states together with the near-threshold $\psi(3770)$, we find that the BESII $R$-value line shape can be well reproduced over the full energy range while retaining the lower-mass $\psi(4160)$ assignment. The enhancement around $4.19$~GeV then arises from the coherent interplay among the nearby $\psi(4040)$, $\psi(4160)$, and $\psi(4220)$ amplitudes, rather than requiring an upward shift of the $\psi(4160)$ mass itself. The additional higher states also naturally describe the line-shape structure in the $4.4$~GeV region. We further show that the seven-resonance coherent amplitude contains six complex zeros, yielding $2^6=64$ mathematically equivalent solutions with identical line shapes but substantially different di-electron widths and relative phases. Comparing these solutions with available experimental information and representative unquenched charmonium predictions, we provide a qualitative assessment of their phenomenological consistency and highlight several solutions that appear more compatible with present information.
		
	\end{abstract}
	
	\maketitle
	
	\section{Introduction}

	As indicated in a recent review article~\cite{Bai:2026atm}, the study of hadron spectroscopy is entering a high-precision era, characterized by the unquenched effect that has been increasingly recognized and emphasized over the past years. This development is instrumental in deepening our understanding of the nonperturbative behavior of the strong interaction, which remains a core issue in modern particle physics.
	
	Historically, the spectroscopy of charmonium was long interpreted within the quenched potential model, which successfully described the low-lying states through a simple $c\bar{c}$ picture~\cite{Eichten:1974af,Eichten:1979ms}. However, this framework began to show severe limitations above the open-charm thresholds. 
	The pivotal turning point arrived in the early 2000s with the high-luminosity $B$-factory experiments (Belle and BaBar), which observed several anomalous resonant structures in the 4.0--4.5~GeV mass region---most notably the $Y(4260)$ and $Y(4360)$~\cite{BaBar:2005hhc,BaBar:2006ait,Belle:2007umv}. 
	These states exhibited properties that could not be reconciled with naive $c\bar{c}$ assignments or conventional potential predictions~\cite{Eichten:1974af,Eichten:1979ms}. 
	This discrepancy triggered a paradigm shift: it became imperative to move beyond the quenched approximation. Consequently, the systematic inclusion of coupled-channel dynamics---i.e., the unquenched effect, where the bare quark core couples strongly to nearby open-flavor meson-meson continua---was recognized as essential for reproducing the observed mass shifts, widths, and line shapes. 
	This historical progression highlights that precise spectroscopy in this energy window demands treating bound-state and scattering states on an equal footing, thus establishing the 4.0--4.5~GeV range as a unique laboratory for probing strong-interaction dynamics beyond the static potential picture.

	The mass parameters and di-lepton widths of the $\psi(4160)$ have already attracted attention in studies based on unquenched charmonium spectroscopy~\cite{Man:2025vmm,Peng:2024blp,Li:2009zu}.
	Here, instead of discussing these quantities solely from the spectroscopy side, we directly revisit the experimental procedure from which they are extracted, namely the inclusive $R$-value analysis. Motivated by the characteristic vector-charmonium spectrum of the unquenched picture, we reanalyze the BESII $R$-value data by including the additional higher vector states expected in this spectrum.

	In the conventional quenched potential-model picture, $\psi(4160)$ is usually assigned predominantly to the $2D$ charmonium state, and its resonance parameters and di-lepton width have historically been extracted mainly from inclusive $R$-value measurements. One longstanding puzzle concerns its mass. Early measurements placed the $\psi(4160)$ near $4.16$~GeV~\cite{DASP:1978dns}, whereas later $R$-scan analyses obtained a value around $4.19$~GeV~\cite{BES:2007zwq}, corresponding to an upward shift of about $30$~MeV. In Ref.~\cite{Peng:2024blp}, we showed, using $B^+\to K^+\mu^+\mu^-$ as an illustrative process, that this apparent shift can be alleviated when the characteristic unquenched charmonium spectrum is taken into account. This observation motivates a direct reassessment of the $R$-value line shape.
	
	A complementary clue is provided by the leptonic coupling of the $\psi(4160)$. Under lepton universality, the di-electron and di-muon widths probe the same electromagnetic coupling, $\Gamma_{e^+e^-}\simeq\Gamma_{\mu^+\mu^-}$. The inclusive $R$-value analysis gives $\Gamma_{e^+e^-}^{\psi(4160)}=0.48\pm0.22$~keV~\cite{BES:2007zwq}, whereas the BESIII analysis of $e^+e^-\to\mu^+\mu^-$ yields, for a representative solution, $\Gamma_{\mu^+\mu^-}^{\psi(4160)}=2.45\pm1.24\pm0.94$~keV~\cite{Ablikim:2020jrn}. Although the latter result has a sizable uncertainty, the substantially different values indicate that the extracted electromagnetic coupling may depend sensitively on the resonance content assumed in the analysis.
	
	This point is especially relevant because the two analyses treat the $4.2$~GeV region differently. The historical $R$-value analysis was based mainly on the conventional quenched spectrum with three states $\psi(4040)$, $\psi(4160)$, and $\psi(4415)$ in 4.0--4.5~GeV range, whereas the BESIII $e^+e^-\to\mu^+\mu^-$ analysis requires an additional structure near $4.22$~GeV in all acceptable solutions~\cite{Ablikim:2020jrn, Farrar:2023zmj}. For a representative solution, this structure, denoted as $S(4220)$, has $M=4216.7\pm8.9\pm4.1$~MeV and $\Gamma=47.2\pm22.8\pm10.5$~MeV. These values are close to the  parameters associated with the $\psi(4220)$ in the characteristic unquenched charmonium spectrum, $M = 4204\sim4243$~MeV and $\Gamma = 26$~MeV.
    The appearance of an additional vector structure near the $\psi(4160)$ region therefore suggests that the conventional three--state resonance content may be insufficient for describing the observed line shapes.
	
	In the characteristic unquenched spectrum, the $\psi(4160)$ is accompanied by a nearby $\psi(4220)$, while additional $\psi(4380)$ and $\psi(4500)$ states appear in the higher-energy region together with the established $\psi(4415)$~\cite{Wang:2019mhs}. The enhancement around $4.16$--$4.22$~GeV may thus reflect the coherent interplay of several nearby vector amplitudes rather than a single isolated $\psi(4160)$ resonance. 
	This possibility provides the central motivation of the present work: we reanalyze the BESII $R$-value data using the characteristic six-state vector-charmonium spectrum and examine whether the structure near $4.19$~GeV can be reproduced while retaining a lower-mass $\psi(4160)$ consistent with quark-model expectations.
	The $\psi(3770)$ is additionally included to describe the lower-energy threshold region. We also investigate the asymmetric enhancement around $4.4$~GeV and the discrete ambiguity associated with the coherent seven-resonance amplitude.
	
	In principle, a simultaneous analysis of the inclusive $R$-value and $e^+e^-\to\mu^+\mu^-$ measurements would provide complementary constraints on the resonance parameters and leptonic couplings of the vector charmonia. At present, however, the available di-muon data are not sufficiently precise and dense over the full energy region for a fully constrained global analysis.
    We therefore use the existing di-muon measurements as complementary evidence for additional resonance content in the $4.2$~GeV region. The phenomenological plausibility of the equivalent solutions obtained from the $R$-value fit is then assessed mainly through comparison with available di-electron widths and representative theoretical calculations.
    
    Future high-precision measurements of both the $R$-value and the di-muon channel at BESIII would enable a more stringent combined determination of the resonance parameters and di-lepton widths of the higher vector charmonium states.
	
    This paper is organized as follows. In Sec.~\ref{II}, we introduce the vector-charmonium spectrum adopted in the analysis and describe the framework used to reanalyze the BESII $R$-value data. We then discuss the discrete ambiguity of the coherent seven-resonance amplitude and the resulting equivalent solutions. These solutions are assessed using available experimental information and representative theoretical expectations, from which a comparatively favored solution is selected to illustrate the fitted line shape. In Sec.~\ref{sec:summary}, we summarize our main results and discuss their implications. The numerical parameters and qualitative assessment of all numerically distinguishable solutions are collected in Appendix~\ref{app}.
	
	\section{Reanalysis of the $R$-value data}\label{II}
	
	\subsection{Quenched vs unquenched charmonium mass spectrum}
	\label{sec:unquenched}
	
	A key issue in the interpretation of the higher vector-charmonium region is the spectroscopy framework adopted in the analysis. In the conventional quenched potential-model picture, the 4.0--4.5~GeV region is mainly described by three established vector charmonia, $\psi(4040)$, $\psi(4160)$, and $\psi(4415)$~\cite{Eichten:1974af,Eichten:1979ms}, which also constituted the conventional resonance content used in early analyses of the inclusive $R$-value data. However, as experimental measurements have entered the era of high-precision hadron spectroscopy, an increasing number of vector structures have been observed in this energy region, posing a growing challenge to the conventional three--state charmonium picture and motivating a more complete description that incorporates unquenched effects.
	
	When unquenched effects are taken into account, the higher vector-charmonium spectrum becomes considerably richer, with six states in 4.0--4.5~GeV. In the framework developed in Refs.~\cite{Wang:2019mhs,Man:2025zfu,Man:2025vmm}, the coupling of the $c\bar c$ configurations to open-charm channels modifies the higher charmonium level structure and leads to a characteristic six-state vector spectrum in the same energy region, consisting of $\psi(4040)$, $\psi(4160)$, $\psi(4220)$, $\psi(4380)$, $\psi(4415)$, and $\psi(4500)$. These states can be organized into three $S$--$D$ pairs, namely the $3S$--$2D$ pair $\psi(4040)$--$\psi(4160)$, the $4S$--$3D$ pair $\psi(4220)$--$\psi(4380)$, and the $5S$--$4D$ pair $\psi(4415)$--$\psi(4500)$. This characteristic spectrum has subsequently been employed in studies of several processes, including $e^+e^-\to\psi(2S)\pi^+\pi^-$~\cite{Wang:2019mhs}, $e^+e^-\to K^+K^-J/\psi$~\cite{Wang:2022jxj}, $e^+e^-\to\pi^+D^0D^{*-}$~\cite{Wang:2023zxj}, $e^+e^-\to\eta J/\psi$~\cite{Peng:2024xui}, and $B^+\to K^+\mu^+\mu^-$~\cite{Peng:2024blp}, indicating that this unquenched resonance content provides a useful phenomenological framework for describing the complex line shapes in the higher-charmonium region.

	It should be emphasized that the term unquenched vector-charmonium spectrum in the present work refers primarily to this characteristic state content and level organization, rather than to a unique set of precisely predicted resonance parameters. Different implementations of unquenched dynamics, including screened-potential, coupled-channel, and phenomenological $S$--$D$ mixing approaches, may yield somewhat different masses, strong decay widths, and di-electron widths for individual states. Such variations are expected for highly excited charmonia, whose properties can be sensitive to open-charm coupled channels and mixing dynamics. Our purpose is therefore not to test a particular numerical realization of an unquenched model, but to examine how the additional resonance content changes the interpretation of the inclusive $R$-value line shape.

	Motivated by this difference in resonance content between the quenched and unquenched descriptions, we revisit the BESII $R$-value data using the characteristic six-state vector spectrum. The $\psi(3770)$ is additionally included to account for the lower-energy resonance contribution in the fitted energy range.
	The masses and widths for highly excited charmonia in this analysis are summarized in Table~\ref{tab:states}.
	These values are either taken from experimental measurements or correspond to  values employed in previous spectroscopic and phenomenological studies.
		
		
	In the present analysis, the resonance parameters are fixed, whereas the di-electron widths $\Gamma_{e^+e^-}^{(j)}$ and the relative phases are determined directly from the $R$-value data. This strategy reduces correlations among the fit parameters and, more importantly, allows us to isolate the effect of the enlarged resonance content and the associated interference pattern. The central question is whether the structure around $4.19$~GeV can be reproduced through the coherent contributions of the nearby $\psi(4040)$, $\psi(4160)$, and $\psi(4220)$ amplitudes while retaining the lower-mass $\psi(4160)$, rather than requiring an upward shift of its mass parameter.
    At higher energies, the additional $\psi(4380)$ and $\psi(4500)$ amplitudes likewise provide a natural framework for examining the asymmetric structure around $4.4$~GeV. Available theoretical calculations and  experimental measurements are used subsequently as complementary information to assess the phenomenological plausibility of the mathematically equivalent solutions obtained from the coherent fit.

    \begin{table*}[htpb]
		\begin{ruledtabular}
			\caption{Masses, widths, and spectroscopic assignments of the vector charmonium states included in the present analysis.
				For the well-established $\psi(3770)$ and $\psi(4040)$, the adopted parameters are taken from the PDG averages~\cite{ParticleDataGroup:2024cfk}.
				For $\psi(4160)$ and $\psi(4415)$, we adopt experimental determinations that are compatible with the corresponding mass and width expectations of the unquenched charmonium spectrum~\cite{DASP:1978dns,Siegrist:1976br}.
				The parameters of $\psi(4220)$ are taken from the BESIII measurement~\cite{BESIII:2016bnd}.
                For the additional higher states $\psi(4380)$ and $\psi(4500)$, the adopted masses and representative total widths are taken from the unquenched charmonium study of Ref.~\cite{Wang:2019mhs}.}
			\label{tab:states}
			\centering
			\begin{tabular}{cccc}
				States & Assignment & Mass (MeV) & Width (MeV)  \\
				\hline
				$\psi(3770)$ &$\psi''_{2S-1D}$  &$3773.7\pm0.7$~\cite{ParticleDataGroup:2024cfk} 		&$27.2\pm1.0$~\cite{ParticleDataGroup:2024cfk} \\
				
				$\psi(4040)$ &$\psi'_{3S-2D}$ &	 $4040\pm4$~\cite{ParticleDataGroup:2024cfk}    &$84\pm12$~\cite{ParticleDataGroup:2024cfk} \\
				
				$\psi(4160)$ &$\psi''_{3S-2D}$ &			$4159\pm{20}$~\cite{DASP:1978dns} &			$78\pm{20}$~\cite{DASP:1978dns} \\
				
				$\psi(4220)$ &$\psi'_{4S-3D}$ & $4222.0\pm{3.1}\pm{1.4}$~\cite{BESIII:2016bnd}& $44.1\pm{4.3}\pm{2.0}$~\cite{BESIII:2016bnd} \\
				
				$\psi(4380)$ &$\psi''_{4S-3D}$ &			$4389 \pm{25} $~\cite{Wang:2019mhs} &			$80$~\cite{Wang:2019mhs} \\
				
				$\psi(4415)$ & $\psi'_{5S-4D}$ &			$4414\pm{7}$~\cite{Siegrist:1976br} &			$33\pm{10}$~\cite{Siegrist:1976br} \\
				
				$\psi(4500)$ & $\psi''_{5S-4D}$ &			$4509 \pm{20}$~\cite{Wang:2019mhs} &			$50$~\cite{Wang:2019mhs} \\
			\end{tabular}
		\end{ruledtabular}
	\end{table*}

	\subsection{Analysis framework}
	\label{sec:formalism}

	The $R$-value is one of the most fundamental observables in $e^+e^-$ annihilation. It is defined as the ratio of the inclusive hadronic production cross section to the lowest-order QED cross section for muon-pair production,
	\begin{equation}
		R(s)=
		\frac{\sigma(e^+e^-\to {\rm hadrons})}
		{\sigma_0(e^+e^-\to\mu^+\mu^-)}
		\label{eq:Rvalue}
	\end{equation}
with
	\begin{equation}
		\sigma_0(e^+e^-\to\mu^+\mu^-)
		=\frac{4\pi\alpha^2}{3s}.
	\end{equation}
	Here,  $\alpha= 1/137$ denotes the electromagnetic fine-structure constant, which characterizes the strength of the electromagnetic interaction. Using this lowest-order QED normalization, the measured $R$-value can be converted into the inclusive hadronic cross section through~\cite{Mo:2010bw}
	\begin{equation}
		\sigma_{\rm had}(s)
		=R(s)\frac{86.85}{s},
		\label{eq:Rtosigma}
	\end{equation}
	where $s$ is the squared center-of-mass energy in units of GeV$^2$.
	
	Following Ref.~\cite{Mo:2010bw}, the smooth nonresonant background contribution is parameterized as
	\begin{equation}
		\sigma_{\rm bkg}(\sqrt{s})
		=a+b(\sqrt{s}-2m_D)-c(\sqrt{s}-2m_D)^2,
		\label{eq:bkg}
	\end{equation}
	where $m_D=1.86966$~GeV, and the parameters $a$, $b$, and $c$ have units of nb, nb/GeV, and nb/GeV$^2$, respectively.
	
	For a vector charmonium state $j$, the resonant contribution is described by the Breit--Wigner amplitude
	\begin{equation}
		A_j(\sqrt{s})=
		\frac{\sqrt{12\pi\,\Gamma_{e^+e^-}^{(j)}\Gamma_j}}
		{s-M_j^2+iM_j\Gamma_j},
		\label{eq:bw}
	\end{equation}
	where $M_j$, $\Gamma_j$, and $\Gamma_{e^+e^-}^{(j)}$ denote the mass, total width, and di-electron width of the state, respectively. Since the total widths of the vector charmonia considered here are dominated by hadronic decays, we approximate $\Gamma_j^{\rm had}\simeq\Gamma_j$.
	
	The resonant amplitudes are added coherently,
	\begin{equation}
		\sigma_{\rm res}(\sqrt{s})=
		\left|
		\sum_{j=1}^{7} A_j(\sqrt{s})e^{i\phi_j}
		\right|^2,
		\label{eq:coherent}
	\end{equation}
	where the subscripts $j=1,2,\cdots,7$ are used to distinguish different amplitudes and phases involving the $\psi(3770)$, $\psi(4040)$, $\psi(4160)$, $\psi(4220)$, $\psi(4380)$, $\psi(4415)$, and $\psi(4500)$, respectively.

	The total hadronic cross section is then written as
	\begin{equation}
		\sigma_{\rm tot}(\sqrt{s})
		=\sigma_{\rm bkg}(\sqrt{s})
		+\sigma_{\rm res}(\sqrt{s}).
		\label{eq:total}
	\end{equation}
	Following Refs.~\cite{BES:2007zwq,Mo:2010bw}, interference between the continuum background and the resonant amplitudes is neglected in the fit. This treatment allows us to focus on the effect of the resonance content and the interference among the higher vector charmonium states while maintaining consistency with previous $R$-value analyses. Since only relative resonance phases are physically relevant, the $\psi(3770)$ amplitude is taken as the phase reference, with $\phi_{1}=0$, and the remaining six relative phases are determined from the fit.

	\subsection{Fit results and discussion}
	\label{sec:fitresults}
	
	Because several resonances contribute coherently to the total line shape, the resonance couplings extracted from the fit are subject to a discrete ambiguity. For seven interfering Breit--Wigner amplitudes, the analytic structure of the amplitude formally generates $2^{7-1}=64$ mathematically equivalent solutions with identical line shapes but different di-electron widths and relative phases~\cite{Bai:2019jrb}. In the present case, two amplitude zeros lie extremely close to the real axis, so that these 64 formal solutions can be organized into 16 numerically distinguishable solutions at the precision relevant here. The mathematical origin of this ambiguity, the corresponding complex zeros, and the parameters of the 16 solutions are presented in Appendix~\ref{app}.
	
\begin{figure}[htbp]
		\centering
		\includegraphics[width=\columnwidth]{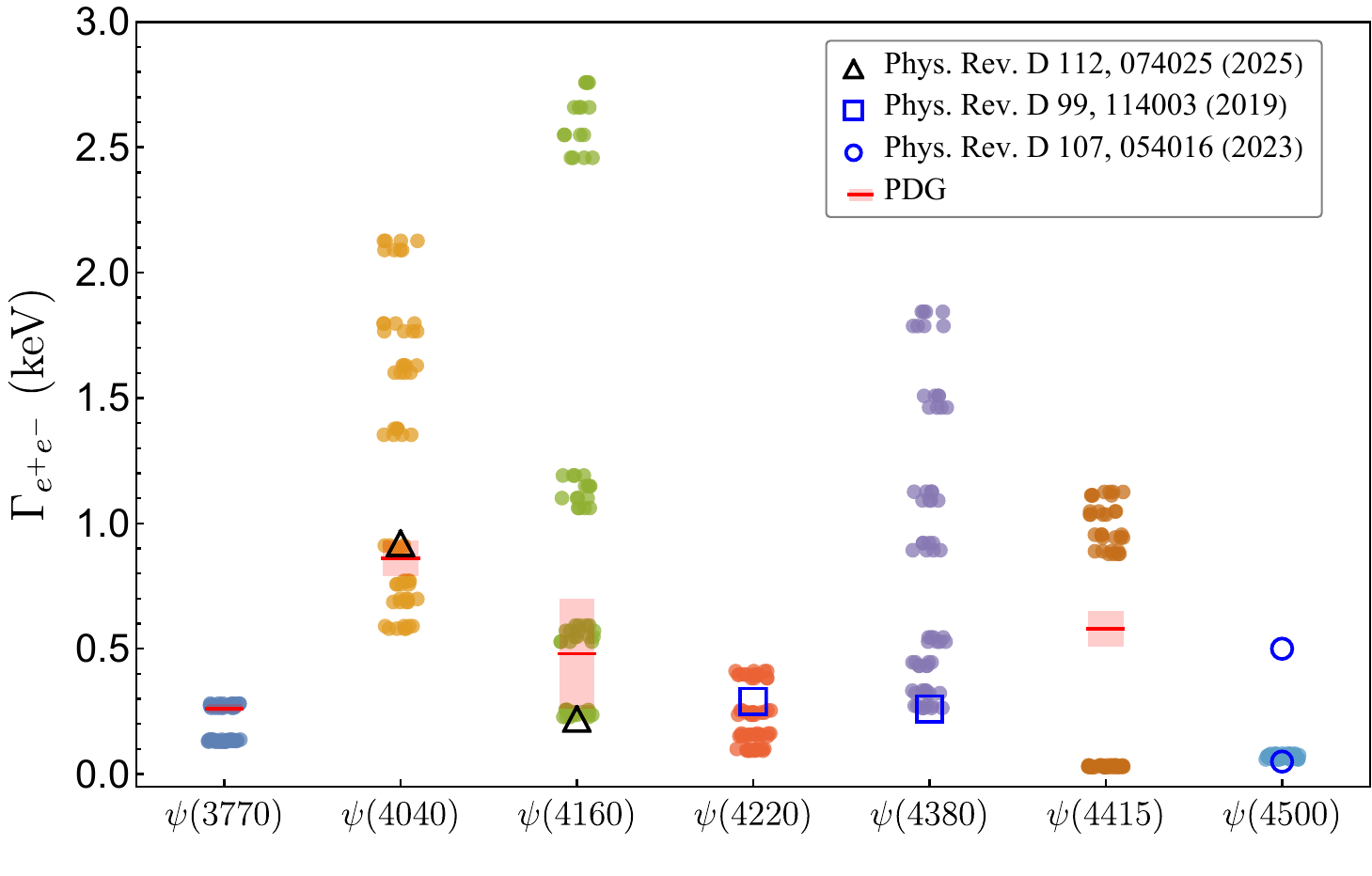}
		\caption{
			Di-electron widths $\Gamma^{(j)}_{e^+e^-}$ of the seven vector charmonium states for the 16 numerically distinguishable solutions of the $R$-value fit. The colored points denote the solutions obtained in the present analysis. The red shaded bands show the PDG reference values and quoted uncertainties for $\psi(3770)$, $\psi(4040)$, $\psi(4160)$, and $\psi(4415)$~\cite{ParticleDataGroup:2024cfk}. The values for $\psi(3770)$ and $\psi(4040)$ are taken as the primary experimental references, whereas those for $\psi(4160)$ and $\psi(4415)$ are shown only for comparison because their previous extractions may be affected by incomplete resonance content in the corresponding energy regions. The open triangles denote the recent coupled-channel predictions for $\psi(4040)$ and $\psi(4160)$~\cite{Man:2025vmm}. The open blue squares denote the theoretical values for $\psi(4220)$ and $\psi(4380)$ from Ref.~\cite{Wang:2019mhs}, while the open blue circles indicate the two alternative predictions for $\psi(4500)$ reported in Ref.~\cite{Wang:2022jxj}.
		}
		\label{fig:widthbranches}
	\end{figure}

	Although these solutions cannot be distinguished by the inclusive $R$-value line shape itself, their fitted di-electron widths exhibit clearly separated discrete patterns, as shown in Fig.~\ref{fig:widthbranches}. The variations are relatively small for $\psi(3770)$ and $\psi(4500)$, whereas substantially larger spreads occur for $\psi(4040)$, $\psi(4160)$, $\psi(4380)$, and $\psi(4415)$.
	
	To assess the phenomenological plausibility of these solutions, we use the available experimental and theoretical information as external guidance. The di-electron widths of the well-established $\psi(3770)$ and $\psi(4040)$ provide the primary experimental references~\cite{ParticleDataGroup:2024cfk}. By contrast, the PDG values for $\psi(4160)$ and $\psi(4415)$ are used only for comparison, since their previous extractions may be affected by incomplete resonance content in the corresponding energy regions. On the theoretical side, the recent coupled-channel calculation with a small $3S$--$2D$ mixing angle provides predictions for the di-electron widths of $\psi(4040)$ and $\psi(4160)$~\cite{Man:2025vmm}, while Refs.~\cite{Wang:2019mhs,Wang:2022jxj} provide representative expectations for $\psi(4220)$, $\psi(4380)$, and $\psi(4500)$.
	
	Based on these external inputs, the 16 solutions can be qualitatively classified according to their overall phenomenological consistency. Solution No.~1 is comparatively favored because it provides the best overall agreement with the established experimental references while remaining broadly compatible with the representative expectations for the higher vector states.
	Solutions No.~2 and No.~4 are marginally compatible. 
	Solution No.~2 gives a reasonable di-electron width for $\psi(3770)$ and several higher-state di-electron widths close to the theoretical expectations, but yields a somewhat smaller $\psi(4040)$ di-electron width than the experimental value. Solution No.~4 reproduces the experimental $\psi(3770)$ and $\psi(4040)$ di-electron widths well, but gives a considerably larger $\psi(4380)$ di-electron width than the representative theoretical expectation.
	
	The remaining solutions are phenomenologically disfavored because they exhibit substantial tension with one or more of the primary experimental references and/or the representative expectations for the higher vector states. For many of these solutions, this tension appears as either a substantially underestimated $\psi(3770)$ di-electron width or an excessively large $\psi(4040)$ di-electron width. The complete assessment is summarized in Appendix~\ref{app}. We emphasize that this classification provides only phenomenological guidance and does not constitute a statistical exclusion, since all 16 solutions reproduce exactly the same $R$-value line shape.
	
	Among these solutions, solution No.~1 shows the best overall consistency with the presently available external information and is therefore chosen as the representative solution for illustrating the fit and the interference pattern. The corresponding background parameters are
	\begin{equation}
		\begin{aligned}
			a &= 13.543\pm0.116~{\rm nb},\\
            b &= 3.556\pm0.218~{\rm nb/GeV},\\
            c &= 3.509\pm0.320~{\rm nb/GeV^2}.
		\end{aligned}
		\label{bkgfit}
	\end{equation}
	The fitted di-electron widths and relative phases are summarized in Table~\ref{tab:fit_parameters}.
	
	\begin{table}[htbp]
 	\begin{ruledtabular}
		\centering
		\caption{Di-electron widths and relative phases of the representative solution No.~1 obtained from the final fit. The $\psi(3770)$ phase is fixed to $\phi_{1}=0$ and taken as the reference.}
		\label{tab:fit_parameters}
		\begin{tabular}{ccc}
			State & $\Gamma_{e^+e^-}^{(j)}$ (keV) & $\phi_j$ (rad) \\
			\midrule
			$\psi(3770)$ & $0.270\pm0.021$ & $0$ (fixed) \\
			$\psi(4040)$ & $0.896\pm0.045$ & $4.459\pm0.085$ \\
			$\psi(4160)$ & $0.571\pm0.041$ & $0.403\pm0.077$ \\
			$\psi(4220)$ & $0.245\pm0.052$ & $3.695\pm0.074$ \\
			$\psi(4380)$ & $0.446\pm0.053$ & $4.113\pm0.169$ \\
			$\psi(4415)$ & $0.033\pm0.036$ & $3.154\pm0.189$ \\
			$\psi(4500)$ & $0.067\pm0.059$ & $1.723\pm0.305$ \\
		\end{tabular}
  	\end{ruledtabular}
	\end{table}
	
	Several features of the representative solution are worth noting.
	The di-electron widths of $\psi(3770)$, $\psi(4040)$, $\psi(4160)$, $\psi(4220)$, and $\psi(4380)$ are relatively well constrained by the fit, indicating that the inclusive $R$-value data retain appreciable sensitivity to their di-electron widths despite the strong interference among the resonances. 
	By contrast, the fitted widths of $\psi(4415)$ and $\psi(4500)$ carry uncertainties comparable to their central values. This indicates that the inclusive data provide only weak independent constraints on the di-electron widths of these states in the strongly overlapping $4.4$~GeV region and these large uncertainties reflect correlations among the nearby states. 
	
	The parametrization contains 17 quantities in total, among which $\phi_{1}=0$ is fixed to define the overall phase convention, leaving 16 free parameters in the fit. The resulting fit over the $3.7$--$4.8$~GeV energy region gives
	\begin{equation}
		\chi^2/{\rm d.o.f.}=0.827.
	\end{equation}
	This value is smaller than $1.05$ reported in the BESII analysis~\cite{BES:2007zwq}, in which, apart from the $\psi(3770)$ contribution, the higher-energy line shape was described using only three vector charmonium states $\psi(4040)$, $\psi(4160)$, and $\psi(4415)$ included in the conventional quenched picture.
	
	\begin{figure}[t]
		\centering
		\includegraphics[width=\columnwidth]{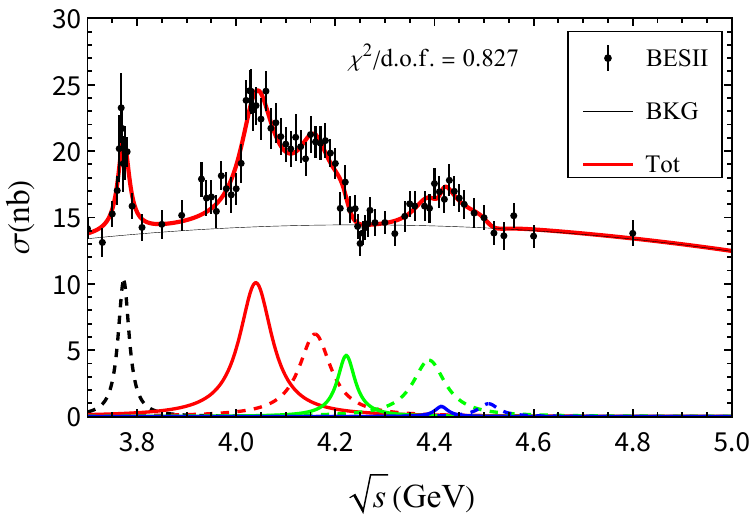}
		\caption{
			Representative fit to the hadronic cross section converted from the BESII $R$-value data~\cite{BES:2001ckj}. The total fit consists of the smooth continuum contribution and the coherent sum of seven vector-charmonium amplitudes. The curves shown in the lower part illustrate the individual resonance contributions.
		}
		\label{fig:fit}
	\end{figure}
	As shown in Fig.~\ref{fig:fit}, the present seven-state analysis provides a good description of the inclusive $R$-value spectrum. Compared with the earlier BESII analysis~\cite{BES:2007zwq}, the enlarged resonance content leads to a substantially different interpretation of the structure around $4.19$~GeV. In the earlier analysis, this enhancement was associated mainly with the $\psi(4160)$, whereas in the present framework it can be reproduced through the coherent interplay among the nearby $\psi(4040)$, $\psi(4160)$, and $\psi(4220)$ amplitudes while retaining a lower $\psi(4160)$ mass.
    The apparent enhancement near $4.19$~GeV therefore need not coincide with the mass parameter of a single resonance, but can instead arise as a multi-resonance line-shape effect.
	
	The presence of an additional vector contribution in the $4.2$~GeV region is also qualitatively consistent with the BESIII analysis of the di-muon process $e^+e^-\to\mu^+\mu^-$, in which an additional structure with $M=4216.7\pm8.9\pm4.1$~MeV and $\Gamma=47.2\pm22.8\pm10.5$~MeV was required to describe the line shape~\cite{Ablikim:2020jrn,Farrar:2023zmj}.
    The historical $R$-value data are thus compatible with the lower-mass $\psi(4160)$ scenario once the enlarged resonance content and the associated interference effects of the characteristic unquenched charmonium spectrum are taken into account.
	
	A similar multi-resonance effect appears in the $4.4$~GeV region. In the quenched spectrum description, the broad and asymmetric enhancement in this region is mainly associated with the $\psi(4415)$. In the present analysis, however, the observed line shape can be reproduced through the coherent contributions of the nearby $\psi(4380)$, $\psi(4415)$, and $\psi(4500)$ amplitudes. This observation may also be relevant to the sizable spread among the experimentally extracted widths of the $\psi(4415)$ in different processes, which range from several tens of MeV to about $110$--$120$~MeV~\cite{ParticleDataGroup:2024cfk}.
    If nearby resonances and their interference are not explicitly included, their contributions may be partially absorbed into the effective mass and width of a single $\psi(4415)$ amplitude. The different apparent line shapes and width determinations reported in various processes may therefore reflect, at least in part, different interference patterns among the overlapping vector states. Consequently, the structure around $4.4$~GeV need not originate from a single isolated $\psi(4415)$ state.
	
	The remaining ambiguity highlights the importance of complementary measurements. After more than 15 years of operation, BESIII has accumulated large data samples in the charmonium energy region, providing an important basis for improving the precision and energy coverage of inclusive $R$-value measurements. More precise $R$-value data may reveal finer line-shape structures and place stronger constraints on the masses, widths, and di-electron widths of the additional vector charmonium states. Independent and more complete measurements of the di-muon process $e^+e^-\to\mu^+\mu^-$ over the same energy region would provide complementary information on the di-electron widths and may help distinguish among the presently equivalent solutions. Together with improved theoretical calculations, such measurements could substantially reduce the remaining ambiguity and lead to a more complete understanding of highly excited charmonium spectroscopy.

	\section{Summary}	\label{sec:summary}

	In summary, we have revisited the BESII $R$-value data within the framework	of unquenched vector-charmonium spectroscopy.
	The main motivation for this study comes from the long-standing puzzle that previous $R$-scan analyses tended to place the mass of the $\psi(4160)$ close to $4.19$~GeV, about $30$~MeV higher than the early experimental measurements around	$4.16$~GeV. We believe that this discrepancy is related to the incomplete understanding of unquenched effects in the spectroscopy of highly excited charmonium states in earlier analyses.
	In particular, the conventional quenched picture contains only three vector charmonium states in the 4.0--4.5~GeV region, whereas the unquenched spectrum contains six states in the same energy range.
	The development of unquenched charmonium spectroscopy over the past two decades has been comprehensively reviewed in Ref.~\cite{Bai:2026atm}. The applicability of this characteristic unquenched spectrum has also been tested in a variety of processes in previous studies~\cite{Wang:2019mhs,Wang:2022jxj,Wang:2023zxj,Peng:2024xui,Peng:2024blp}.
	
	Based on the unquenched vector-charmonium spectrum, we include the contributions from $\psi(3770)$, $\psi(4040)$, $\psi(4160)$, $\psi(4220)$, $\psi(4380)$, $\psi(4415)$, and $\psi(4500)$ in a coherent analysis of the BESII $R$-value data~\cite{BES:2001ckj}.
	A representative solution provides a good description of the measured line shape while keeping the $\psi(4160)$ mass fixed at $4.159$~GeV.
	This result shows that, once the unquenched spectrum is taken into account, the inclusive $R$-value data themselves do not require the $\psi(4160)$ mass near $4.19$~GeV. 
	The historically observed upward shift can instead be understood as a line-shape effect arising	mainly from the interference among the nearby $\psi(4040)$, $\psi(4160)$, and $\psi(4220)$ amplitudes, rather than as direct evidence that the mass of the $\psi(4160)$ must be close to $4.19$~GeV.
	This result demonstrates that the historical $R$-value data are compatible with a lower-mass $\psi(4160)$ scenario once the richer resonance content expected in the characteristic unquenched spectrum and the corresponding interference effects are taken into account.

	We have further investigated the multiple-solution structure of the seven-resonance coherent amplitude. The six complex zeros formally generate $2^{7-1}=64$ mathematically equivalent solutions. Because two zeros lie extremely close to the real axis, these solutions can be organized into 16 numerically distinguishable branches at the precision relevant here. These solutions generate the same resonant intensity along the physical energy axis, while their di-electron widths and relative phases can differ substantially. This ambiguity implies that the $R$-value data alone are insufficient to uniquely determine the di-electron widths of the individual vector charmonium states.
	To provide additional phenomenological guidance, we compare the fitted $\Gamma^{(j)}_{e^+e^-}$ values with available external information. In particular, the PDG values for the well-established $\psi(3770)$ and $\psi(4040)$ states are used as the primary experimental references~\cite{ParticleDataGroup:2024cfk}. The recent coupled-channel predictions for $\psi(4040)$ and $\psi(4160)$ with a small $3S$--$2D$ mixing angle~\cite{Man:2025vmm}, together with representative theoretical expectations for $\psi(4220)$, $\psi(4380)$, and $\psi(4500)$ from Refs.~\cite{Wang:2019mhs,Wang:2022jxj}, are used as complementary theoretical references.
	This comparison allows several solutions with better overall consistency to be identified as phenomenologically favored or suggested, although the discrete ambiguity cannot be removed on this basis alone because the theoretical predictions remain model dependent and experimental information on the higher states is still limited.
	
	After more than 15 years of operation, BESIII has accumulated large data samples in the charmonium energy region. Further improvements in the precision and energy coverage of $R$-value measurements may therefore become possible, potentially revealing finer structures in the inclusive line shape and providing clearer constraints on the masses, widths, and di-electron widths of the additional vector charmonium states associated with the unquenched spectrum. More complete and precise measurements of the di-muon channel $e^+e^-\to\mu^+\mu^-$ at BESIII, together with improved theoretical calculations, would provide complementary information on the di-electron widths and help resolve the remaining discrete ambiguities among the equivalent solutions. Such combined information will be important for establishing a more complete and reliable spectroscopy of highly excited charmonium states.
	
	\begin{acknowledgments}
This work is supported by the Natural Science Foundation of Gansu Province (No. 26RCKA012 and No. 25JRRA799), the National Natural Science Foundation of China under Grants No. 12335001 and No. 12247101, the ``111 Center" under Grant No. B20063, the Fundamental Research Funds for the Central Universities (lzujbky-2023-stlt01), and Lanzhou City High-Level Talent Funding.
	\end{acknowledgments}
	
	\bibliography{references.bib}

	\appendix

	\section{Equivalent solutions of the coherent multi-resonance amplitude}
	\label{app}
	
	The multiple solutions of the coherent fit have a simple analytic origin. Introducing the complex pole positions
	\begin{equation}
		p_j=M_j^2-iM_j\Gamma_j,
	\end{equation}
	and absorbing the magnitude and phase of each coupling into a complex residue $c_j$, the seven-resonance amplitude can be written as
	\begin{equation}
		\mathcal A(s)
		=
		\sum_{j=1}^{7}\frac{c_j}{s-p_j}
		=
		\frac{P_6(s)}
		{\displaystyle\prod_{j=1}^{7}(s-p_j)}.
		\label{eq:rational}
	\end{equation}
	The numerator is a sixth-order polynomial,
	\begin{equation}
		P_6(s)=C\prod_{k=1}^{6}(s-z_k),
		\label{eq:numerator}
	\end{equation}
	where $z_k$ denote the six complex zeros of the amplitude.
	
	For real $s$,
	\begin{equation}
		|s-z_k|^2=|s-z_k^\ast|^2.
		\label{eq:zeroconj}
	\end{equation}
	Therefore, replacing any subset of the zeros by their complex conjugates,
	\begin{equation}
		z_k\rightarrow z_k^\ast,
	\end{equation}
	leaves $|\mathcal A(s)|^2$ unchanged on the real axis. Since the continuum background is added incoherently in the present fit, this transformation also leaves the total fitted cross section unchanged. Each transformed polynomial corresponds to a different set of residues $c_j^\prime$, and hence to different di-electron widths $\Gamma_{e^+e^-}^{(j)}$ and relative phases $\phi_j$, while producing exactly the same resonant line shape. For $N$ interfering Breit--Wigner poles, there are generically $2^{N-1}$ mathematically equivalent solutions~\cite{Bai:2019jrb}. For the present seven-resonance amplitude, this gives
	\begin{equation}
		N_{\rm sol}=2^6=64.
	\end{equation}
	
	For the representative solution shown in Fig.~\ref{fig:fit}, the six zeros are
	\begin{align}
		z_1&=14.380-0.540i,\nonumber\\
		z_2&=17.059-0.802i,\nonumber\\
		z_3&=18.044-0.745i,\nonumber\\
		z_4&=18.163-1.332\times10^{-6}i,\nonumber\\
		z_5&=19.436-0.132i,\nonumber\\
		z_6&=20.537-1.190\times10^{-4}i
		\qquad ({\rm GeV}^2).
		\label{eq:zeros}
	\end{align}
	
\begin{figure}[htbp]
		\centering
		\includegraphics[width=\columnwidth]{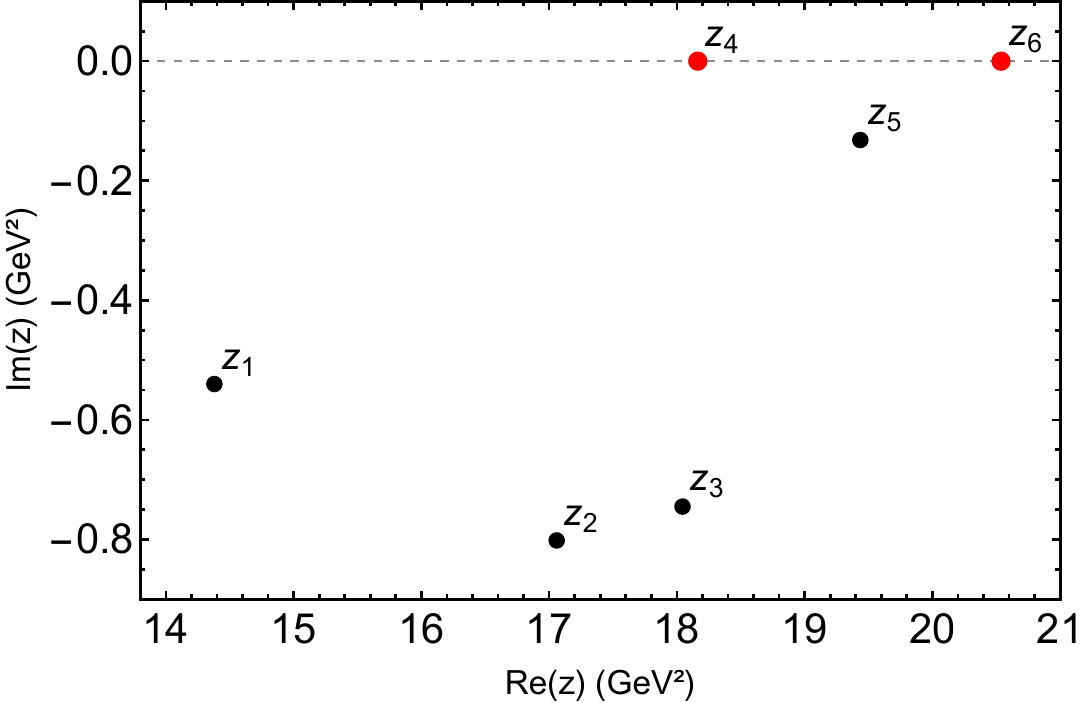}
		\caption{Zeros of the numerator of the seven-resonance amplitude in the complex $s$ plane. The two red points, corresponding to $z_4$ and $z_6$, lie extremely close to the real axis and therefore generate numerically almost indistinguishable solutions under complex conjugation.}
		\label{fig:zeros}
	\end{figure}

	These zeros are displayed in Fig.~\ref{fig:zeros}. In particular, $z_4$ and $z_6$ lie extremely close to the real axis. Conjugating either of these two zeros therefore induces only negligible numerical changes in the corresponding amplitude parameters. Consequently, although the six complex zeros formally generate $2^6=64$ mathematically equivalent solutions, the solutions associated with the conjugation of $z_4$ and $z_6$ are numerically almost indistinguishable at the precision relevant to the present analysis. The 64 formal solutions can therefore be organized into $\frac{64}{2^2}=16$ numerically distinguishable solutions. We have also verified directly that all zero-conjugated solutions reproduce numerically identical line shapes over the entire fitted energy region. The 16 numerically distinguishable solutions are summarized in Table~\ref{tab:16solutions}. Although they reproduce the same $R$-value line shape, they yield substantially different di-electron widths and relative phases.

 	The qualitative assessment in Table~\ref{tab:16solutions} is based primarily on the comparison of the fitted di-electron widths with the experimentally better-established values for $\psi(3770)$ and $\psi(4040)$~\cite{ParticleDataGroup:2024cfk}. The PDG values for $\psi(4160)$ and $\psi(4415)$ are included only as additional references, since their previous extractions may be affected by incomplete resonance content in the corresponding energy regions. Representative theoretical predictions for $\psi(4040)$ and $\psi(4160)$ are taken from Ref.~\cite{Man:2025vmm}, while those for $\psi(4220)$, $\psi(4380)$, and $\psi(4500)$ are taken from Refs.~\cite{Wang:2019mhs,Wang:2022jxj}. The symbols in the last column therefore provide only a qualitative indication of the phenomenological consistency of the different solutions.
	
	\begin{table*}[htbp]
 	\begin{ruledtabular}
		\centering
		\caption{Parameters of the 16 numerically distinguishable solutions of the coherent seven-resonance fit. The first seven numerical columns give the di-electron widths $\Gamma^{(j)}_{e^+e^-}$ in keV, while the following six columns give the relative phases in radians. The phase $\phi_{1}=0$ is fixed as the reference. The symbols $\checkmark$, $\triangle$, and $\times$ denote, respectively, comparatively favored, marginally compatible, and phenomenologically disfavored solutions according to a qualitative comparison with the presently available experimental and theoretical information.}
		\label{tab:16solutions}
		\resizebox{\textwidth}{!}{
			\begin{tabular}{c|ccccccc|cccccc|c}
				No.
				& $\Gamma_{e^+e^-}^{(1)}$
				& $\Gamma_{e^+e^-}^{(2)}$
				& $\Gamma_{e^+e^-}^{(3)}$
				& $\Gamma_{e^+e^-}^{(4)}$
				& $\Gamma_{e^+e^-}^{(5)}$
				& $\Gamma_{e^+e^-}^{(6)}$
				& $\Gamma_{e^+e^-}^{(7)}$
				& $\phi_{2}$
				& $\phi_{3}$
				& $\phi_{4}$
				& $\phi_{5}$
				& $\phi_{6}$
				& $\phi_{7}$
				& Assessment \\
				\midrule
				1  & 0.270 & 0.896 & 0.571 & 0.245 & 0.446 & 0.033 & 0.067 & 4.459 & 0.403 & 3.695 & 4.113 & 3.154 & 1.722 & $\checkmark$ \\[1pt]
				2  & 0.265 & 0.686 & 0.247 & 0.097 & 0.272 & 0.028 & 0.059 & 4.051 & 5.598 & 1.543 & 5.544 & 4.489 & 2.733 & $\triangle$ \\[1pt]
				3  & 0.265 & 0.698 & 0.256 & 0.100 & 0.921 & 0.889 & 0.068 & 4.084 & 5.668 & 1.654 & 5.817 & 3.318 & 2.406 & $\times$ \\[1pt]
				4  & 0.271 & 0.912 & 0.592 & 0.254 & 1.508 & 1.048 & 0.076 & 4.492 & 0.473 & 3.805 & 4.386 & 1.983 & 1.396 & $\triangle$ \\[1pt]
				5  & 0.275 & 1.601 & 1.148 & 0.156 & 0.333 & 0.030 & 0.063 & 5.061 & 2.575 & 5.686 & 4.308 & 3.300 & 1.702 & $\times$ \\[1pt]
				6  & 0.275 & 1.629 & 1.191 & 0.162 & 1.126 & 0.955 & 0.072 & 5.093 & 2.645 & 5.796 & 4.581 & 2.128 & 1.375 & $\times$ \\[1pt]
				7  & 0.281 & 2.090 & 2.659 & 0.396 & 0.545 & 0.035 & 0.071 & 5.468 & 3.663 & 1.554 & 2.877 & 1.965 & 0.691 & $\times$ \\[1pt]
				8  & 0.281 & 2.127 & 2.758 & 0.410 & 1.844 & 1.125 & 0.081 & 5.501 & 3.733 & 1.665 & 3.150 & 0.794 & 0.365 & $\times$ \\[1pt]
				9  & 0.129 & 0.580 & 0.228 & 0.093 & 0.264 & 0.027 & 0.058 & 0.901 & 2.280 & 4.456 & 2.083 & 1.020 & 5.517 & $\times$ \\[1pt]
				10 & 0.130 & 0.590 & 0.236 & 0.097 & 0.893 & 0.878 & 0.067 & 0.934 & 2.349 & 4.567 & 2.356 & 6.132 & 5.190 & $\times$ \\[1pt]
				11 & 0.132 & 0.757 & 0.528 & 0.237 & 0.432 & 0.032 & 0.066 & 1.309 & 3.368 & 0.324 & 0.652 & 5.968 & 4.506 & $\times$ \\[1pt]
				12 & 0.132 & 0.770 & 0.548 & 0.246 & 1.462 & 1.035 & 0.075 & 1.342 & 3.437 & 0.435 & 0.925 & 4.797 & 4.179 & $\times$ \\[1pt]
				13 & 0.134 & 1.353 & 1.061 & 0.151 & 0.323 & 0.029 & 0.062 & 1.911 & 5.540 & 2.315 & 0.847 & 6.113 & 4.485 & $\times$ \\[1pt]
				14 & 0.135 & 1.377 & 1.101 & 0.157 & 1.091 & 0.943 & 0.071 & 1.944 & 5.610 & 2.426 & 1.120 & 4.942 & 4.159 & $\times$ \\[1pt]
				15 & 0.137 & 1.766 & 2.458 & 0.383 & 0.528 & 0.035 & 0.070 & 2.319 & 0.345 & 4.467 & 5.699 & 4.779 & 3.475 & $\times$ \\[1pt]
				16 & 0.137 & 1.797 & 2.549 & 0.397 & 1.787 & 1.112 & 0.080 & 2.352 & 0.415 & 4.578 & 5.972 & 3.607 & 3.148 & $\times$ \\
		\end{tabular}}
  	\end{ruledtabular}
	\end{table*}

\end{document}